\documentclass[aps,prl,twocolumn,superscriptaddress]{revtex4-2}
\usepackage{graphicx}
\usepackage{bm}
\usepackage{amsmath, amssymb}
\usepackage{comment}
\usepackage[normalem]{ulem}
\usepackage{lipsum}
\usepackage{appendix}
\usepackage{physics}
\usepackage{xcolor}
\usepackage{appendix}
\usepackage{mathtools}
\usepackage[breaklinks=true,colorlinks,citecolor=blue,linkcolor=blue,urlcolor=blue]{hyperref}
\usepackage{subfigure}
\usepackage[T1]{fontenc}
\usepackage[all]{hypcap}
\usepackage{bbm}

\newcommand{\figref}[2]{\hyperref[#1]{\autoref*{#1}#2}}
\newcommand{\aref}[1]{\hyperref[#1]{App.~\ref*{#1}}}

\newcommand{\HXXZ}[0]{H_\mathrm{XXZ}}

\newcommand{\Sc}[0]{\mathcal{S}}

\newcommand{\Sztot}[0]{S^z_\mathrm{tot}}

\begin{document}

\title{Temperature and integrability-breaking correspondence via adiabatic transformations}
\author{Hyeongjin Kim}
\email{hkim12@bu.edu}
\altaffiliation{equal contribution}
\affiliation{Department of Physics, Boston University, Boston, Massachusetts 02215, USA}

\author{Souvik Bandyopadhyay}
\email{sbandyop@bu.edu}
\altaffiliation{equal contribution}
\affiliation{Department of Physics, Boston University, Boston, Massachusetts 02215, USA}

\author{Anatoli Polkovnikov}
\affiliation{Department of Physics, Boston University, Boston, Massachusetts 02215, USA}

\date{\today}

\begin{abstract}
We reveal a correspondence between temperature and integrability-breaking in classical and quantum many-body systems through the lens of geometry and adiabatic transformations. Decreasing the temperature, obtained in a standard way through the derivative of entropy with respect to energy, steers the system towards an integrable point despite strong integrability-breaking interactions. Auto-correlation functions of local observables exhibit slow relaxation dynamics, which violates ergodicity on the approach to this integrable point. 
Subsequently, the average fidelity susceptibility of stationary states satisfies scaling relations near the integrable point, in close analogy with continuous phase transitions. We further find that the dynamical exponent encompassing relaxation can be different in the quantum and classical models, depending on dimension of the systems. Collectively, our results establish temperature as a tunable control parameter for chaos and puts it on equal footing with integrability-breaking perturbations. 
\end{abstract}

\maketitle

\noindent \textbf{\textit{Introduction.}}\textemdash The distinction between integrable and chaotic dynamics is fundamental to understanding emergent statistical mechanics in isolated systems~\cite{rigol2008thermalization,polkovnikov2011nonequilibrium}. Integrable systems preclude thermalization due to an extensive set of conserved quantities that constrain the dynamics. In contrast, typical chaotic Hamiltonian systems thermalize to states dictated by conserved total energy.

In quantum mechanics, thermalization is commonly understood through the spectral statistics of energy eigenstates and operator matrix elements (see Refs.~\cite{Michael_Berry_1989,Haake1991,stockmann_1999,borgonovi2016quantum,d2016quantum,Deutsch:2018rev} for review). Matrix elements of a typical local observable in the spectrum of a chaotic Hamiltonian satisfy the eigenstate thermalization hypothesis (ETH)~\cite{Srednicki_1999,Deutsch:2018rev,d2016quantum}. On the other hand, classical chaos is traditionally understood through exponential trajectory sensitivity to initial conditions~\cite{oseledets_1968,Ott_2002,strogatz2018nonlinear}. Ergodicity is determined by comparing the relaxation of local observables to their equilibrium values~\cite{rigol2008thermalization,polkovnikov2011nonequilibrium}. 

However, weakly-nonintegrable systems can display long-lived non-ergodic dynamics. In classical systems, chaos in this regime is confined to narrow resonant islands according to the Kolmogorov-Arnold-Moser (KAM) theorem~\cite{kolmogorov_1954,arnold_1963,moser_1962,Chierchia:2010}. In near-integrable quantum systems, large off-diagonal matrix elements between nearby energy states break ergodicity: rare resonances leading to slow relaxation. Although such behavior is expected to be transient in the thermodynamic limit, it often dominates experimentally relevant time scales.
\begin{figure}[!t]
    \includegraphics[width=\columnwidth]{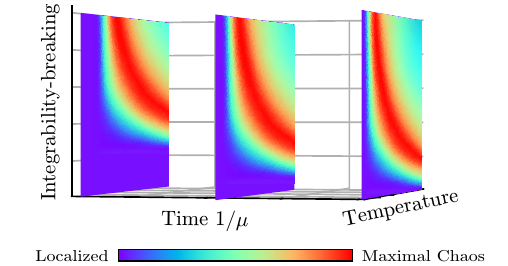}
	\caption{\emph{Temperature and integrability-breaking correspondence.} A schematic illustration of the correspondence parameterized by time $1/\mu$, the effective temperature $T$ as defined in \autoref{eq:temperature_definition}, and integrability-breaking perturbation strength.}
	\label{fig:main}
\end{figure}

At low temperatures (or in the dilute limit), connectivity in accessible state space is suppressed despite strong interactions. Fewer resonances prevent extensive mixing of states, leading to integrable-like behavior. In this paper, we explore this correspondence between temperature and integrability-breaking in both classical and quantum systems, from the viewpoint of geometry and adiabatic transformations.

As our main quantitative diagnostics of chaos, we use the \emph{adiabatic gauge potential} (AGP)~\cite{kolodrubetz2017geometry,berry2009transitionless,adiabatic_population,pandey2020adiabatic}. 
In quantum mechanics, the AGP generates adiabatic deformations of eigenstates under infinitesimal perturbations~\cite{pandey2020adiabatic}, while, in classical mechanics, it is the generator of trajectory-preserving canonical transformations and of deformed conservation laws~\cite{Patra_2017,kim2025defining}. 
The AGP accurately detects long-time dynamical instabilities, quantifying the sensitivity of states/trajectories to perturbations: it captures mechanisms underlying chaotic behavior of both quantum~\cite{pandey2020adiabatic,leblond2021universality} and classical systems~\cite{kim2025defining,KarveRoseCampbell2025,karve2025diffusionsignaturechaos,vanovac2026weakintegrabilitybreakingperturbations}. The variance of the AGP is the fidelity susceptibility $\chi$. More generally, for a multi-component coupling space, the covariance matrix of the AGP is the quantum geometric tensor, quantifying the complexity of adiabatic transformations. The fidelity susceptibility and the geometric tensor diverge in the thermodynamic limit (see also Ref.~\cite{jarzynski1995geometric}) but can be regularized with a small frequency cutoff $\mu$. 
Similar to the behavior of ground state susceptibility near phase transitions~\cite{zanardi2006ground,venuti2007quantum,kolodrubetz2013classifying,sierant2019fidelity}, $\chi$ is very sensitive to perturbations that lead to ergodicity-breaking, especially along integrable directions ~\cite{leblond2021universality,bulchandani2022onset,pandey2020adiabatic,sels2020dynamical,surace2023weak,orlov2023adiabatic,correale2023probing,kim2023integrability,vanovac2024finitesize,bhattacharjeeSharpDetectionOnset2024a,swietek2025scaling,swietek2025fading}.

Our main result is visually illustrated in \autoref{fig:main}, which showcases the relationship between the cutoff time $1/\mu$, temperature $T$, and the integrability-breaking perturbation strength in a many-body system. While \autoref{fig:main} is illustrative, it is based on actual numerical analysis done in this work. The color code shows the logarithm of the regularized fidelity susceptibility. The horizontal axis shows the slices of different waiting times $1/\mu$. For the classical case study, we analytically prove that temperature renormalizes the effective nonlinearity around a harmonic integrable limit, while the low-temperature quantum model can be characterized by a nearly integrable dilute gas of well-defined local quasiparticles. 

Whether we consider weak integrability-breaking at high temperature or strong integrability-breaking at low temperature, the integrable and ergodic regions are separated by an intermediate KAM-like chaotic non-ergodic regime (maximal chaos)~\cite{sels2020dynamical,leblond2021universality,Skvortsov2022PRB,swietek2025fading,kim2025defining}. With increasing cutoff time $1/\mu$, both quantum and classical systems behave chaotically for lower temperatures and smaller integrability breaking perturbation strengths.

\noindent \textbf{\textit{Models}}\textemdash We study the one-dimensional XXZ spin chain with or without incommensurate potential, which we call disorder. In the usual quantum description for spin-$\tfrac{1}{2}$ particles (with $\hbar = 1$ and periodic boundary conditions),
\begin{align}\label{eq:xxz}
\begin{split}
    \HXXZ &= -\Big[\frac{J}{2}\sum^{L-1}_{j=0}\left(\hat{S}_j^+ \hat{S}_{j+1}^- + \hat{S}_j^- \hat{S}_{j+1}^+ \right) + \Delta \sum^{L-1}_{j=0} \hat{S}_j^z \hat{S}_{j+1}^z\\
    &\quad \quad \quad + \Delta^\prime \sum^{L-1}_{j=0} \hat{S}_j^z \hat{S}_{j+2}^z + W \sum^{L-1}_{j=0} \cos(2\pi \kappa j) \hat{S}^z_j \Big],
\end{split}
\end{align}
where $\hat{S}_j^\pm = S_j^x \pm i S_j^y$, $J$ sets the coupling strength, $\Delta$ is the nearest-neighbor interaction strength, $\Delta^\prime$ is the next-nearest-neighbor interaction strength, $W$ is the disorder strength, and the spacing $\kappa = (\sqrt{5} -1)/2$ is incommensurate to the lattice, which induces localization of eigenstates similar to disordered spin chains~\cite{falcao2024mbl}. For the quantum model with $W \neq 0$, we use open boundary conditions. $\HXXZ$ is integrable when $\Delta^\prime = 0$, and also at $W \to \infty$ where eigenstates are localized due to the strong disorder. The interaction $\Delta^{\prime}$ favors delocalization by opening up many-body resonance channels between free fermion states. Therefore,  either $W^{-1}$ or $\Delta^{\prime}$ plays the role of the integrability-breaking perturbation strength. Further, the model conserves the total $z$-magnetization $\Sztot\equiv\expval*{\sum_j\hat{S}_j^z}$.

In the classical model, we remove the hats and replace each spin with a vector $\vb{S}_j$: $\norm{\vb{S}_j}^2 = (S_j^x)^2 + (S_j^y)^2 + (S_j^z)^2 = 1$. Additionally, we replace the commutators with Poisson brackets: $\poissonbracket*{S_j^\alpha}{S_k^\beta} = \delta_{jk} \epsilon_{\alpha \beta \gamma} S^\gamma_j$, where $\alpha,\beta \in \{x,y,z\}$, $\delta_{jk}$ is the Kronecker delta, and $\epsilon_{\alpha \beta \gamma}$ is the Levi-Civita symbol.

\noindent \textbf{\textit{Effective temperature.}}\textemdash
Quantum mechanically, we uniformly average over a fixed magnetization sector $\Sztot=M=(2\rho-1)L/2$ to probe different entropic regions of the Hilbert space, where \(\rho\) is the \emph{magnetic density}. Similarly, in the classical model, we impose the constraint \(S^z_{\rm tot}  = M = (2\rho-1)L\) on phase-space averages, targeting specific entropic regions. However, as entropy itself is defined only up to an additive constant in finite classical systems and becomes ill-defined in the limit \(\rho \to 0\), we define the \textit{effective temperature} \(T\):
\begin{equation}
\label{eq:temperature_definition}
    \frac{1}{T(\rho)} = \frac{\Delta \mathcal{S}(\rho)}{\Delta E(\rho)} .
\end{equation}
Here, \(\mathcal{S}(\rho)\) and \(E(\rho)\) denote the entropy and mean energy, respectively, at fixed $\rho$; 
\(\Delta \mathcal{S}(\rho)\) and \(\Delta E(\rho)\) are the differences in entropy and mean energy between the sectors with \(\Sztot = M\) and \(\Sztot = M+1\) in the quantum model and $S^z_{\rm tot}=M$ and $S^z_{\rm tot}=M+\delta M$ in the classical model. The reason we use the effective and not the physical temperature is that it is difficult to work with low-energy states, especially in quantum systems, due to their low density. We therefore consider ensembles in which all energy eigenstates within a given magnetization sector are equally probable. Consequently, we can study significantly larger system sizes, keeping the Hilbert-space dimension fixed. Then $T(\rho)$ captures the rate of change of the density of accessible states with energy, similar to standard temperature. If we consider the grand-canonical ensemble, where $n_\uparrow = \rho L$ is the total number of spin-up particles, the effective temperature in the quantum case, for example, is given by $T(\rho)=(\Delta+\Delta')\frac{\tanh(\beta_\mathrm{phys}\mu/2)}{\beta_\mathrm{phys}\mu}$, where $\beta_{\rm phys}=1/T_\mathrm{phys}$ is the inverse physical temperature and $\mu$ is the chemical potential. In the low-temperature regime of interest, $T(\rho)$ is proportional to $T_\mathrm{phys}$.

In the quantum model, $\Sc(\rho)=\log\Omega(\rho,L)$, where $\Omega(\rho,L)$ is the number of states. In the thermodynamic limit, $\Sc(\rho)/L\approx-\rho\log\rho-(1-\rho)\log(1-\rho)$. Similarly, the energy is calculated as $E(\rho) = \Omega(\rho,L)^{-1} {\rm Tr}_{M}[H_{\rm XXZ}]$. Following standard algebra [see Supplemental Material (SM)~\cite{SupplementalMaterial}], 
\begin{equation} \label{eq:temperature_quantum}
    T(\rho) = \frac{(\Delta + \Delta^{\prime})(1-2\rho)}{\log[(1-\rho)/\rho]},
\end{equation}
which is maximal at $\rho=\tfrac{1}{2}$: $T(\rho=\tfrac{1}{2}) = (\Delta + \Delta^\prime)/2$.

In the classical model, $T(\rho)$ is set by the Lagrange multiplier $\lambda(\rho)$ from the constraint $S^z_\mathrm{tot}=M$ (see SM~\cite{SupplementalMaterial}):
\begin{equation} \label{eq:temperature_classical}
    T(\rho) = \frac{2(\Delta + \Delta^{\prime})(1-2\rho)}{\lambda(\rho)},
\end{equation}
such that $\lambda^{-1}-\coth{\lambda} = 2\rho-1$. This effective temperature is maximal at $\rho=\tfrac{1}{2}$: $T(\rho=\tfrac{1}{2}) =  2(\Delta+\Delta^\prime)/3$. Both the classical and quantum effective temperatures have the limits $T(0)=T(1)=0$ and satisfy $T(\rho)=T(1-\rho)$.

\noindent \textbf{\textit{Classical model.}}\textemdash The correspondence between temperature and integrability-breaking can be established exactly in the classical model. Unlike the quantum model, the classical model at $W = 0$ is not integrable for either non-zero $\Delta$ or $\Delta^\prime$. Consider the Holstein-Primakoff representation~\cite{holstein1940field,Auerbach1994Interacting}: for spin on the $j$-th site,
\begin{align} \label{eq:HP}
    S^z_j = n_j- 1,~S^+_j = a^\ast_j\sqrt{2-n_j},~S^-_j = \sqrt{2-n_j}a_j,
\end{align}
with $\poissonbracket{a_i}{a^\ast_j}=-i\delta_{ij}$ and $n_j=a_j^\ast a_j$. The magnetic density maps to $\rho=\frac{1}{2L}\sum_{j}n_j$. We expand around $n_j=0,~\forall j$, or, equivalently, $S_j^z = -1$ for all spins and find (see SM~\cite{SupplementalMaterial})
\begin{equation}
\HXXZ = H^{(0)} + H^{(2)} + \mathcal{O}(n^3),
\end{equation}
such that $H^{(0)}$ has $\mathcal{O}(n)$ terms, and
\begin{align}
\begin{split}
H^{(2)} &= -\sum^{L-1}_{j=0} \Big[\Delta\, n_j n_{j+1} + \Delta^\prime \, n_j n_{j+2}  \\
& \quad \quad \quad \quad  -\frac{J}{4} (n_j + n_{j+1})(a^\ast_j a_{j+1} + a_j a^\ast_{j+1}) \Big]
\end{split}
\end{align}
is nonlinear, contributing to the chaoticity.

Thereafter, we rewrite the Hamiltonian in terms of rescaled variables $a_j= \sqrt{\rho}\alpha_j$, satisfying the equations of motion,
\begin{equation}
    i \dot a_j=\frac{\partial H}{\partial a_j^\ast}\quad \implies \quad i\dot \alpha_j=\frac{1}{\rho} \frac{\partial H}{\partial \alpha_j^\ast}.
\end{equation}
The equations for $\alpha_j$ are still Hamiltonian, provided we replace $H^{(0)}(\{a_j,a_j^\ast\})+H^{(2)}(\{a_j,a_j^\ast\})$ with $H^{(0)}(\{\alpha_j,\alpha_j^\ast\})+\rho H^{(2)}(\{\alpha_j,\alpha_j^\ast\})$. This emergent Hamiltonian can be interpreted as a general nonintegrable model with $H=H_0+\epsilon V$, where $H_0=H^{(0)}$ is the integrable part and $\epsilon V=\epsilon H^{(2)}$ is the integrability-breaking perturbation. We see that $\rho$ (or $T(\rho)$) plays the role of $\epsilon$. Higher order terms in the expansion lead to sub-leading contributions at low temperatures. Thus, at low temperatures, the spins are almost polarized along the $z$-axis and spin wave excitations are nearly harmonic, leading to suppressed Lyapunov exponents (see SM~\cite{SupplementalMaterial} and Ref.~\cite{ruidas2026manybodychaosemergespresence}). In SM~\cite{SupplementalMaterial}, we find that the classical model exhibits confined chaos~\cite{milani_an_1992,milani_stable_1997,laskar_chaotic_2008,winter_short_2010,goldfriend2019equilibration,goldfriend2020quasi,pereira_confined_2024,kim2025confineddeconfinedchaosclassical}, where the Lyapunov timescale is asymptotically faster than the relaxation timescale with respect to $\rho$.

\noindent \textbf{\textit{Spectral function and fidelity susceptibility.}}\textemdash To probe the temperature and integrability-breaking correspondence while keeping classical and quantum mechanics on an equal footing, we study the response of a local, traceless in each magnetization sector, observable $O \to O-\langle O \rangle$ ($\langle O \rangle=\Omega^{-1}\rm{Tr}[O]\cdot\mathbbm{1}$ in quantum and $\langle O \rangle$ is the phase-space average in classical). 

For the quantum model, the spectral function $\Phi(\omega)$ is defined (see Ref.~\cite{d2016quantum} for review) as
\begin{equation}\label{eq:spec_def}
    \Phi(\omega) = \frac{1}{2\Omega(\rho,L)}\int^\infty_{-\infty} \frac{\dd{t}}{2\pi} e^{i\omega t} \sum_n \mel{n}{\poissonbracket{O(t)}{O(0)}_+}{n},
\end{equation}
where $\ket{n}$ are eigenstates of the Hamiltonian, $\{A,B \}_+$ is the anti-commutator, and $\int_{-\infty}^{\infty}\Phi(\omega^{\prime}) \dd \omega^{\prime} = \norm{O}^2$. We calculate the normalized spectral function $\Phi(\omega)/\norm{O}^2$, which we henceforth denote by $\Phi(\omega)$. For the classical model,  the trace is replaced by a phase space integral over the spin degrees of freedom at a fixed magnetic density.

The fidelity susceptibility $\chi$ is given as
\begin{equation} \label{eq:fidelity}
    \chi(\mu) = \int^\infty_{-\infty} \dd{\omega} \frac{\omega^2}{(\omega^2 + \mu^2)^2} \Phi(\omega),
\end{equation}
where $\mu > 0$ is a regularizer (frequency cutoff). $\chi$ is equivalent to the Hilbert-Schmidt norm of the AGP, where $\mu^{-1}$ acts as a finite time cutoff in the Källén-Lehmann representation ~\cite{berry1994chaoticlassical,jarzynski1995geometric,manjarres2023adiabaticdrivingparalleltransport,Sugiura_2021}. Approaching the limit $\mu\rightarrow0^+$ is equivalent to resolving smaller energy scales (longer time scales), which are generally limited by the Heisenberg scale $\omega_\mathrm{H}$ in quantum systems: un-regularized typical $\chi$ in quantum systems corresponds to $\mu\sim \omega_\mathrm{H} \sim \exp[-\mathcal S]$. For classical systems, a natural choice is $\mu \sim C/t_\mathrm{evo}$, where $t_\mathrm{evo}$ is the evolution time and $C \geq 1$ is a constant.

$\chi(\mu)$ serves as a diagnostic for chaos and integrability through its scaling with respect to $\mu$ as $\mu \to 0^+$. As $\chi(\mu) \sim \Phi(\mu) / \mu$ for sufficiently small $\mu$, the scaling of $\Phi(\omega)$ at low frequencies encodes the same information. While these behaviors generally depend on the observable~\cite{pandey2020adiabatic,kim2023integrability}, we focus on the case where $O$ is an integrability-preserving perturbation: $O = \partial_\Delta \HXXZ$, unless otherwise stated. Refs.~\cite{pandey2020adiabatic,leblond2021universality} analyzed various cases and found two distinct behaviors:
\begin{equation}\label{eq:chi_spec_regimes}
    \begin{array}{lll}
        \mbox{Ergodic/ETH}, &\chi \sim 1/\mu, & \Phi(\omega \to 0) \sim C>0,\\
        \mbox{Integrable}, &\chi \sim \log(1/\mu), & \Phi(\omega \to 0) \to 0.\\
    \end{array}
\end{equation}
 In low-dimensional chaotic~\cite{kim2025defining} and disordered models~\cite{sels2020dynamical, Skvortsov2022PRB,Iwanek2024Universal}, the ergodic and integrable regions are separated by an intermediate chaotic non-thermalizing regime where $\Phi(\omega) \sim 1/\omega^{1-1/z}$ and $\chi(\mu) \sim 1/\mu^{2-1/z}$ with $z>1$. In many-body systems, the intermediate regime with $\chi\sim 1/\mu^2$ (maximal chaos)~\cite{sels2020dynamical,leblond2021universality,Skvortsov2022PRB,swietek2025fading,kim2025defining} emerges when $\mu \sim \Gamma$ with $\Gamma$ denoting the relaxation rate~\cite{leblond2021universality,swietek2025scaling,kim2025confineddeconfinedchaosclassical}.
 
\begin{figure}[!t]
    \includegraphics[width=\columnwidth]{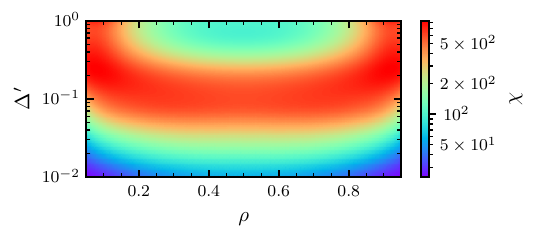}
	\caption{\emph{Correspondence between magnetic density $\rho$ and integrability-breaking $\Delta^\prime$ in the clean XXZ model (quantum).} The fidelity susceptibility $\chi$ at different densities shows that the low density (temperature) states are closer to integrability. Parameters: $J=\sqrt{2}$, $\Delta=(\sqrt{5}+1)/4$, $W=0$ in $\HXXZ$, $\mu = 4\times 10^{-3}$, and in the semi-momentum sector $k = (2\pi/L) \left\lfloor L/4 \right\rfloor$ for $L\in [22, 80]$.}
	\label{fig:clean_xxz_phase}
\end{figure}

\noindent \textbf{\textit{Quantum model.}}\textemdash First, we demonstrate that the maxima of $\chi$, at a fixed frequency cutoff $\mu$, accurately detects the integrable-chaos transition with both Hamiltonian interactions and temperature. \autoref{fig:clean_xxz_phase} is a density plot of $\chi$ with $\Delta^{\prime}$ and $\rho$, for a fixed frequency cutoff $\mu\gg\omega_\mathrm{H}$ at $W=0$. 
The maximum $\chi$ peak reaches a minimum $\Delta^{\prime}$ near $\rho\approx \tfrac{1}{2}$ corresponding to the largest temperature. This drift marks a dynamical crossover between the ergodic and integrable regimes: a correspondence between magnetic density and integrability-breaking strength required for thermalization.

Numerically, we find (see SM~\cite{SupplementalMaterial}) that high $T$ and low $\Delta^{\prime}$ lead to $\Phi(\omega)\propto\omega^{-2}$ tail of the spectral function. This behavior is consistent with the exponential decay governed by Fermi's Golden Rule (FGR) and is standard for ergodic systems~\cite{kim2025confineddeconfinedchaosclassical}. 
However, at low $T$ and high $\Delta^{\prime}$, the spectral function develops a $\omega^{-1}$ tail near the transition. This indicates the emergence of an anomalously slow, non-perturbative regime separating integrable and ergodic regions, which is reminiscent of critical slowing down near continuous phase transitions. This behavior has also been found in interacting disordered systems~\cite{KravtsovAltshuler2018, sels2020dynamical, Iwanek2024Universal} and in low-dimensional classical chaotic models within the mixed phase-space regime~\cite{kim2025defining}. This asymptotic behavior of the spectral function is broadly known as $1/f$-noise~\cite{Paladino_2014}. In our study, we believe that the emergence of this intermediate regime results from two-body collisions between quasi-particles preserving integrability in one-dimensional systems; relaxation is a higher-order effect. Moreover, the $1/\omega$ scaling indicates that the Thouless time is exponentially long in inverse perturbation strength~\cite{sels2020dynamical}: the model remains integrable in any finite order of perturbation theory (see SM~\cite{SupplementalMaterial} for complementary studies in a quantum XXZ ladder model).

\begin{figure}[!t]
    \includegraphics[width=\columnwidth]{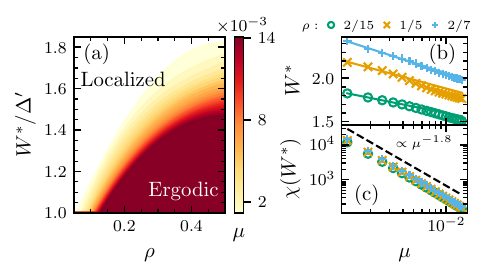}
	\caption{\emph{Onset of localization at different temperatures in the disordered quantum model.} (a): Rescaled susceptibility peak position, $W^\ast/\Delta^\prime$, against density $\rho$.
    (b): Drift of $W^{*} \propto -\log\mu$ as one approaches $\omega_\mathrm{H}$ by decreasing $\mu$. (c): Maximum susceptibility scales as $\chi(W^\ast)\propto1/\mu^{1.8}$. Parameters: $J=2.0$, $\Delta=1.0$, $\Delta^{\prime}=1.5$ in $\HXXZ$, and $L \in [18, 36]$. $\mu \gtrsim \omega_\mathrm{H}^{{\rm max}}$, where $\omega_\mathrm{H}^{\mathrm{max}}$ is the maximum Heisenberg frequency across all values of $\rho$.} 
	\label{fig:onset_localization}
\end{figure}

With non-zero disorder, the transition between ergodic and localized regime is defined by a similar dynamical boundary in the $W$-$\rho$ plane at a fixed  $\Delta^{\prime}$. The susceptibility peak located at $W=W^\ast$ separates the localized regime ($W^{-1} \to 0$) and the delocalized ergodic regime ($W^{-1} \to \infty$). \figref{fig:onset_localization}{a} showcases that the onset of localization drifts towards higher disorder strengths as temperature increases. At lower temperatures, the available Hilbert space is much smaller and localization can persist at much smaller disorder strengths. In Refs.~\cite{LuitzLaflorencieAlet2015MBLedge} and \cite{Guo2021}, similar results were interpreted as a drift of the critical disorder for many-body localization transition with respect to energy. However, later works showed that, at least in the studied regime, there is no transition~\cite{sels2020dynamical, Sels2021Markovian, MorningstarKhemani2022} and thus the boundary indicates a frequency-cutoff dependent crossover that is similar to the one we found here using the maximum of $\chi$.

\begin{figure}[!htbp]
    \includegraphics[width=\columnwidth]{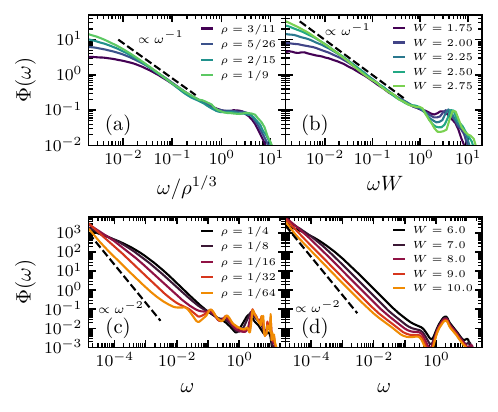}
	\caption{\emph{Spectral function $\Phi(\omega)$ in the disordered XXZ model.} (a)-(b) refer to quantum while (c)-(d) refer to the classical model. (a): At fixed $W = 1.5$, the spectral function scales as $\Phi(\omega) \propto \rho^{1/3}/\omega$ as $\rho\to 0$. (b): At fixed $\rho = 1/2$, the spectral function scales as $\Phi(\omega) \propto (\omega W)^{-1}$ as $W \to \infty$. While not shown, $\omega_\mathrm{H} \gtrsim 4 \times 10^{-4}$ in (a) and $\omega_\mathrm{H} \gtrsim 7 \times 10^{-4}$ in (b). Further, $\omega_\mathrm{Th} \approx 6\times 10^{-4} \text{ to } 4\times 10^{-5} $ in (a) and $\omega_\mathrm{Th} \approx 5 \times 10^{-4} \text{ to } 3 \times 10^{-6}$ in (b) from high $\rho$ or $W^{-1}$ to low $\rho$ or $W^{-1}$.
    In (c)-(d), $\Phi(\omega) \propto \omega^{-2}$, indicating exponential relaxation dynamics, when approaching integrability by varying either $W$ or $\rho$. In (c), $W = 5$ and, in (d), $\rho = 1/2$ are fixed. Parameters: For (a)-(b), $J=2$, $\Delta=1$, $\Delta^{\prime}=1.5$, and $L\in [18,36]$. For (c)-(d), $J=\sqrt{2}$, $\Delta=(\sqrt{5}+1)/4$, $\Delta^{\prime}=1$, and $L = 320$.} 
	\label{fig:combined_spectral}
\end{figure}

The boundary separating the localized/integrable and ergodic regimes is driven by the formation of a low-frequency peak in the spectral function as in Refs.~\cite{leblond2021universality,bulchandani2022onset,pandey2020adiabatic,sels2020dynamical,surace2023weak,orlov2023adiabatic,correale2023probing,kim2023integrability,bhattacharjeeSharpDetectionOnset2024a,swietek2025scaling,swietek2025fading}. Near the transition, we numerically extract scaling relations $\Phi(\omega) \propto \rho^{1/3}/\omega$ at a fixed $W$ and $\Phi(\omega) \propto 1/(W \omega)$ at a fixed temperature, respectively in \figref{fig:combined_spectral}{a} and \figref{fig:combined_spectral}{b}. 
Extrapolating this $1/\omega$ scaling to $\omega \to 0$ in the thermodynamic limit requires a low-frequency cutoff where this growth stops in order to preserve the sum rule $\int \Phi(\omega) \dd\omega=1$: 
i.e., Thouless frequency scales as $\omega_\mathrm{Th} \propto \exp(-\rho^{-1/3})$ at fixed $W$ and $\omega_\mathrm{Th} \propto \exp(-W)$ at fixed $\rho$ (see SM~\cite{SupplementalMaterial} and Refs.~\cite{sels2020dynamical, Iwanek2024Universal} for a related analysis).
This indicates that the relaxation time becomes exponentially long when $\rho \to 0$ or $W \to \infty$. We find that $\omega_{\rm Th}\ll\omega_{\rm H}$ (for most cases), and so the system does not thermalize for accessible system sizes shown in \autoref{fig:combined_spectral}. At constant $\rho$, increasing $W^{-1}$ or $\Delta^\prime$ pushes the system into the ergodic limit, leading to constant spectral weights at low frequencies as in \autoref{eq:chi_spec_regimes}.

The dynamic nature of this transition is characterized by the dependence on the frequency cutoff $\mu$. At a fixed density, $W^\ast$ drifts (see \figref{fig:onset_localization}{b}) as $W^\ast \propto - \rho^{1/3} \log{\mu}$ (see SM~\cite{SupplementalMaterial}), implying that thermalization time increases as $\rho$ decreases. This scaling is consistent with the earlier asymptotic behavior of the Thouless frequency, as the maximum of $\chi$ is expected to be at $\mu \sim \omega_{\rm Th}$ ~\footnote{Although this scaling is compatible with the asymptotic scaling of $\omega_{\rm Th}$ with $W$ and $\rho$ separately, we lack sufficient data to confirm the full scaling function $\omega_{\rm Th}(\rho,W)$.}.
\figref{fig:onset_localization}{c} shows that $ \chi(W^\ast) \propto \mu^{-1.8}$ across all magnetic densities, consistent with the $1/\omega$ tail of the spectral function.

\noindent \textbf{\textit{Quantum-classical correspondence.}}\textemdash Both quantum and classical models display qualitatively similar features on approaching the integrable limit by either reducing the integrability-breaking strength ($W^{-1} \to 0$) or density ($\rho \to 0$) [temperature ($T \to 0$)]. 

This similarity is observed in \figref{fig:combined_spectral}{a} and \figref{fig:combined_spectral}{b} for quantum and \figref{fig:combined_spectral}{c} and \figref{fig:combined_spectral}{d} for classical model, where we show the spectral function $\Phi(\omega)$ while varying either $\rho$ or $W$. We observe distinct signatures of integrability as $\rho \to 0$ or $W^{-1} \to 0$: the onset frequency of the low-frequency power-law tail systematically shifts towards $\omega = 0$. This indicates that, as the system becomes more integrable, the characteristic frequencies of its internal dynamics become slower until eventually the spectral weight at low frequency vanishes as in \autoref{eq:chi_spec_regimes}. In SM~\cite{SupplementalMaterial}, we analyze the relaxation of spin current and see similar slowing down of spin transport near integrability, which can be experimentally verified.

While the qualitative shift of the spectral weight towards low frequencies is similar, the relaxation dynamics reveal a fundamental distinction between the classical and quantum models. At low frequencies, the classical spectral function scales as $\omega^{-2}$, consistent with FGR. On the other hand, the quantum model exhibits an $\omega^{-1}$ tail, violating FGR and indicating logarithmically slow relaxation. As discussed, this difference is likely due to quasi-particles behaving as hardcore gas for the quantum model in the dilute limit, which maps to free fermions and is integrable in one dimension. Therefore, integrability in the quantum model is much more robust, leading to longer lifetimes. For a quantum model with ladder geometry, the spectral function at low temperature approaches the $1/\omega^2$ scaling as in the classical case (see SM~\cite{SupplementalMaterial}). Conversely, the $1/\omega$ scaling of the spectral function for a different but related classical model was observed at strong disorder~\cite{Iwanek2024Universal}. Therefore, we believe that the scaling of spectral function in the crossover region separating integrable and ergodic regimes depends on the nature of the integrability-breaking.

\noindent \textbf{\textit{Conclusion.}}\textemdash By studying the stability of adiabatic transformations, we demonstrate a correspondence between temperature and integrability-breaking perturbation strength. The drift of maximum fidelity susceptibility reveals that the stability of the ergodic phase and thermalization time is deeply linked to temperature, leading to specific scaling forms. In summary, we believe that the \emph{temperature and integrability-breaking correspondence} is generic: chaotic and integrable dynamical regimes are determined by the interplay between the strength of the ``kick'' (perturbation) and temperature setting the volume of the ``room'' (density of states) that the system can explore during its evolution.

One further direction is to explore the onset of chaos and universality at low temperatures, when the spectrum has gapless excitations. The transition between ergodic and localized regimes bears a striking resemblance to continuous phase transitions~\cite{zanardi2006ground,venuti2007quantum,kolodrubetz2013classifying,sierant2019fidelity}. One can anticipate the emergence of different universality classes of thermalization either at weak integrability-breaking or low temperature. Another direction is to develop various perturbative expansions in temperature as an integrability-breaking perturbation and to construct approximate conservation laws, connecting with existing constructions in nearly integrable models~\cite{birkhoff1927,orlov2023adiabatic,pawlowski2025LIOM}.

\noindent \textbf{\textit{Note}}\textemdash While preparing the manuscript, we came across a recent work~\cite{ruidas2026manybodychaosemergespresence} on the low-temperature suppression of decorrelator growth in a classical Heisenberg magnet, which is consistent with our Lyapunov exponents data.

\noindent \textbf{\textit{Acknowledgements}}\textemdash The authors thank Tongyu Zhou for creating the schematic illustration. We acknowledge fruitful discussions with  Monika Aidelsburger, Sourav Bhattacharjee, Marin Bukov, Michael Flynn, Markus Heyl, Chris Laumann, Maciej Lewenstein, Dries Sels, and Lev Vidmar. S.B. thanks ICFO Barcelona for hospitality where a part of this study was performed. This work was supported by the grants NSF DMR-2412542 and AFOSR FA9550-21-1-0342. The authors acknowledge that the computational work in this paper was performed on the Shared Computing Cluster administered by the Boston University Research Computing Services. Numerical computations for quantum systems were performed using \texttt{QuSpin}~\cite{Quspin1,Quspin2}, while those for classical systems were done using \texttt{DynamicalSystems.jl}~\cite{Datseris2018,DatserisParlitz2022}.

\nocite{benettin1976kolmogorov,toda2012statistical}
\bibliography{refs}

\end{document}


\title{Supplemental material for ``Temperature and integrability-breaking correspondence via adiabatic transformations''}
\author{Hyeongjin Kim}
\affiliation{Department of Physics, Boston University, 590 Commonwealth Avenue, Boston, Massachusetts 02215, USA}

\author{Souvik Bandyopadhyay}
\affiliation{Department of Physics, Boston University, 590 Commonwealth Avenue, Boston, Massachusetts 02215, USA}

\author{Anatoli Polkovnikov}
\affiliation{Department of Physics, Boston University, 590 Commonwealth Avenue, Boston, Massachusetts 02215, USA}
\maketitle

\tableofcontents

\section{Derivation of effective temperature}\label{app:temperature}

Here, we complete the derivation of the effective temperature $T(\rho)$ introduced in the main text. It is defined as the inverse derivative of the entropy with respect to energy, where both are implicit functions of the magnetic density $\rho$.

\subsection{Quantum model}
\begin{figure}[!b]
    \centering
    \includegraphics[width=0.5\columnwidth]{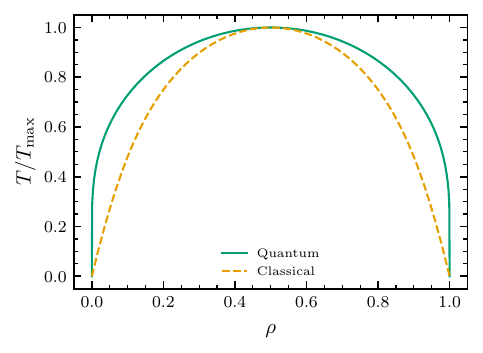}
	\caption{\textit{Effective temperature in the disorder-less quantum and classical XXZ models.} We plot the effective temperature $T$ against $\rho$ using \autoref{eq:temperature_quantum} for the quantum model and \autoref{eq:temperature_classical} for the classical model. We normalize the temperature by their corresponding maximum temperature $T_\mathrm{max} \equiv T(\rho=1/2)$. Parameters: $\Delta = 1$, $\Delta^\prime = 1.5$, and $W = 0$ in $\HXXZ$.}
	\label{fig:temperature}
\end{figure}

In the quantum model, we average uniformly over all states within a fixed magnetization sector. Entropy is then determined by the number of states in that sector. Consider the quantum Hamiltonian defined in the main text,
\begin{equation}
    H_{\rm XXZ} = - \left[ \frac{J}{2}\sum^{L-1}_{j=0}\left(\hat{S}_j^+ \hat{S}_{j+1}^- + \hat{S}_j^- \hat{S}_{j+1}^+ \right) + \Delta \sum^{L-1}_{j=0}\hat{S}_j^z \hat{S}_{j+1}^z + \Delta^\prime \sum^{L-1}_{j=0} \hat{S}_j^z \hat{S}_{j+2}^z + W\sum^{L-1}_{j=0}\cos(2\pi \kappa j) \hat{S}^z_j \right].
\end{equation}
As the total $z$-magnetization $S^z_{\rm tot} = \expval*{\sum_j\hat{S}_j^z}$ is conserved, we can simply fix the density of the up-spins $\rho=n_{\uparrow}/L$ in each magnetization sector. Consequently, the entropy of that sector is given by $\mathcal{S}= \log\Omega(\rho,L)$, where $\Omega(\rho,L)=\begin{pmatrix}L\\ \rho L\end{pmatrix}$. In the thermodynamic limit $L\rightarrow\infty$, at fixed $\rho$, $\mathcal{S}/L\approx -\rho\log\rho-(1-\rho)\log(1-\rho)$ via Stirling's approximation. To define temperature, we further calculate the mean energy as a function of the magnetic density as $E(\rho) = \Omega(\rho,L)^{-1} {\rm Tr}_{M}[H_{\rm XXZ}]$, where the trace is taken over a fixed total magnetization sector $M=\Sztot=(2\rho-1)L/2$. First, note that the hopping ($J$) term does not contribute to the mean energy since it connects different computational basis states and thus has vanishing diagonal matrix elements; its contribution to the uniform trace over a fixed magnetization sector is zero. 
Noting that, at large $L$, different spins are uncorrelated in an infinite-temperature ensemble with $\langle \hat{S}_j^z\rangle=(2\rho-1)/2$, we find~\footnote{In our derivations of quantum and classical temperature, we assume periodic boundary conditions. Note that using open boundary conditions leads to non-extensive differences in our expression of energy $E$ and thus our expression of temperature $T$ remains unchanged in the thermodynamic limit.}
\begin{equation}
    E(\rho) = -\frac{L}{4}(\Delta+\Delta^{\prime})(2\rho-1)^2-\frac{W}{2}(2\rho-1)\frac{\sin{(\pi\kappa L)}\cos{(\pi\kappa(L+1))}}{\sin{(\pi\kappa)}},
\end{equation}
where $\kappa=(\sqrt{5}-1)/2$ is incommensurate with the lattice. Note that the $W$-contribution is non-extensive and can be dropped in the thermodynamic limit. Given these expressions, we find
\begin{equation} \label{eq:temperature_quantum}
    \frac{1}{T(\rho)} = \frac{\partial_\rho \mathcal{S}(\rho)}{\partial_\rho E(\rho)} =\frac{\log\left(\frac{1-\rho}{\rho}\right)}{(\Delta+\Delta^{\prime})(1-2\rho)}.
\end{equation}
The effective temperature definition has a removable singularity at $\rho\rightarrow 1/2$, where we simply define 
\begin{equation}
    T\left(\rho= \frac{1}{2}  \right) \equiv \lim\limits_{\rho\rightarrow \frac{1}{2}} T(\rho) =\frac{\Delta+\Delta^{\prime}}{2}.
\end{equation}
We plot the temperature $T$ against magnetic density $\rho$ in \autoref{fig:temperature}.

\subsection{Classical model}
For the classical model, we follow the same procedure as above and define temperature through the derivative of entropy with respect to energy as we change magnetic density. Suppose that the state of one classical spin is given by the normalized distribution $p(\vb{S})$, where $\vb{S}$ is the vector $(\sin\theta\cos\phi,\sin\theta\sin\phi,\cos\theta)$. Then its entropy $s$ is defined by the following phase space average:
\begin{equation}
    s = -\int \dd\mathcal{V} p(\vb{S})\log p(\vb{S}),
\end{equation}
where $d\mathcal{V}=(4\pi)^{-1}\sin\theta \dd\theta \dd\phi$ is the volume element. Instead of working with the fixed magnetization ensemble, we impose the constraint that the average z-magnetization for the spin is fixed, i.e., $\langle S^z\rangle=m=2\rho-1$.
This can be done using Lagrange's method, leading to the Gibbs-like spin probability distribution $p(\vb{S})$:
\begin{equation}
    p(\vb{S}) = \frac{e^{-\lambda S^z}}{Z(\lambda)},
\end{equation}
where $\lambda$ is the Lagrange multiplier associated with $z$-magnetization and $Z(\lambda) = \int \dd \mathcal{V} e^{-\lambda S^z}$ is the partition function. The partition function can further be calculated analytically as
\begin{equation}
    Z(\lambda) = \frac{1}{4\pi}\int_0^{\pi}\sin\theta e^{-\lambda\cos\theta}\dd\theta\int_0^{2\pi}\dd\phi = \frac{\sinh{\lambda}}{\lambda}.
\end{equation}
The average z-magnetization is then given as
\begin{equation}
    \langle S^z\rangle = -\frac{\partial\log Z(\lambda)}{\partial\lambda}=\frac{1}{\lambda}-\coth{\lambda}=m.
\end{equation}
Similarly, the entropy is given by
\begin{equation}
    s = -\int \dd\mathcal{V} p(\vb{S}) (-\lambda S^z-\log{Z}) = \lambda m + \log{Z}.
\end{equation}
Further, since entropy is extensive, $s=\mathcal{S}/L$ is the entropy density of the system of $L$ spins, where $\mathcal{S}$ is the total entropy. Therefore, the rate of change of entropy with magnetic density is given as
\begin{equation}
    \frac{\partial s}{\partial \rho} = 2\lambda \implies \frac{\partial \mathcal{S}}{\partial \rho} = 2L\lambda.
\end{equation}
Similarly, the total energy of the $S^z_{\rm tot}=mL=(2\rho-1)L$ sector and its rate of change with magnetic density are given by
\begin{equation}
    E(\rho) \approx -L(2\rho-1)^2(\Delta+\Delta^{\prime}) \implies \frac{\partial E}{\partial \rho}= 4L(\Delta + \Delta^{\prime})(1-2\rho),
\end{equation}
where we again neglect the $\mathcal{O}(1)$ contribution of the quasi-periodic disorder. Therefore, we obtain the following definition of effective temperature in the classical model (see \autoref{fig:temperature}):
\begin{equation} \label{eq:temperature_classical}
\frac{1}{T(\rho)} =  \frac{\partial_\rho \mathcal{S(\rho)}}{\partial_\rho E(\rho)} = \frac{\lambda(\rho)}{2(\Delta + \Delta^{\prime})(1-2\rho)},
\end{equation}
where the Lagrange multiplier $\lambda(\rho)$ is set by $\tfrac{1}{\lambda(\rho)}-\coth{\lambda(\rho)}=2\rho-1$, $\lambda(\rho < \tfrac{1}{2}) > 0$, and $\lambda(\rho > \tfrac{1}{2}) < 0$. At $\rho = \tfrac{1}{2}$, this temperature reaches its maximum
\begin{equation}
    T\left(\rho= \frac{1}{2}  \right) \equiv \lim\limits_{\rho\rightarrow \frac{1}{2}} T(\rho) =\frac{2}{3}(\Delta+\Delta^{\prime}).
\end{equation}

\section{Holstein-Primakoff representation}\label{app:holstein-primakoff}
We complement the derivations of how $\rho$ and thus $T(\rho)$ parametrize the non-linearity and chaoticity of the classical $\HXXZ$ model (with $W = 0$) using the Holstein-Primakoff representation:
\begin{align} \label{eq:sm_schwingerboson}
\begin{split}
    S^z_j &= n_j - 1, \\
    S^+_j &= a^\ast_j \sqrt{2-n_j}, \\
    S^-_j &= \sqrt{2-n_j} a_j,
\end{split}
\end{align}
with $\poissonbracket{a_i}{a_j^\ast} = -i\delta_{ij}$ and $n_j = a_j^\ast a_j$. Using $\rho = \frac{1}{2L} \sum_j n_j$ and Taylor expanding \autoref{eq:sm_schwingerboson} near $\rho = 0$ ($n_j = 0,  \forall j$), we have
\begin{align}
    \begin{split}
        S^z_j &= n_j - 1, \\
        S^+_j &\approx \sqrt{2}a_j^\ast \left(1-\frac{n_j}{4}\right) = \sqrt{2}\left(a^\ast_j - \frac{a_j^\ast n_j}{4}\right), \\
        S^- &\approx \sqrt{2}\left(1 - \frac{n_j}{4} \right)a_j = \sqrt{2} \left(a_j - \frac{a_j n_j}{4} \right),
    \end{split}
\end{align}
where we use the approximation $\sqrt{1 - n_j/2} \approx 1 - n_j/4$ for $n_j \approx 0$. Then, the terms in $\HXXZ$ can be broken down into the following:
\begin{align}
\begin{split}
    \frac{1}{2} \left(S_j^+ S_{j+1}^- + \mathrm{h.c.} \right) &= a^\ast_j a_{j+1} + a_j a^\ast_{j+1} - \frac{a_j^\ast a_{j+1} n_{j+1}}{4} - \frac{a^\ast_j n_j a_{j+1}}{4} \\
    & \quad - \frac{a_j a^\ast_{j+1} n_{j+1}}{4} - \frac{a_j n_j a^\ast_{j+1}}{4} + \mathcal{O}(n^3) \\
    &\approx \left(1 - \frac{1}{4} (n_j + n_{j+1}) \right) (a^\ast_j a_{j+1} + a_j a^\ast_{j+1}),
\end{split}
\end{align}
where we drop the $\mathcal{O}(n^3)$ terms as we focus on terms only up to $\mathcal{O}(n^2)$. The $zz$ interaction terms give
\begin{align}
    S_j^z S_{j+1}^z &= (n_j - 1)(n_{j+1} - 1) = 1 - (n_j + n_{j+1}) + n_j n_{j+1}, \\
    S_j^z S_{j+2}^z &= (n_j - 1)(n_{j+2} - 1) = 1 - (n_j + n_{j+2}) + n_j n_{j+2}.
\end{align}
Summing these above equations gives us $\HXXZ = H^{(0)} + H^{(2)} + \mathcal{O}(n^3)$ from the main paper. Note that we ignore the $-L(\Delta + \Delta^\prime)$ term as it only contributes a constant shift in energies: that is, $H^{(0)}$ now only contains the $\mathcal{O}(n)$ terms.

Now, we end this section by proving that $H^{(0)}$ is indeed diagonal in Fourier space:
\begin{equation}
    H^{(0)} = \sum_j [\Delta (n_j + n_{j+1})+ \Delta^\prime (n_j + n_{j+2}) - J(a^\ast_j a_{j+1} + a_j a^\ast_{j+1})].
\end{equation}
By symmetry of the summation (with periodic boundary conditions), we have
\begin{equation}
    H^{(0)} = \sum_j [2(\Delta + \Delta^\prime) n_j - J(a^\ast_j a_{j+1} + a_j a^\ast_{j+1})].
\end{equation}
We take the Fourier transform
\begin{equation}
    a_j = \frac{1}{\sqrt{L}} \sum_k e^{ikj} a_k
\end{equation}
and find that
\begin{equation}
    H^{(0)} = \frac{1}{L} \sum_{k,q} \sum_j \left[2(\Delta + \Delta^\prime) e^{-i(k-q)j} a^\ast_k a_q - J\left(e^{-i(k-q)j} e^{iq} a^\ast_k a_q + e^{-i(k-q)j} e^{-ik} a^\ast_k a_q \right)\right].
\end{equation}
We use the identity $\delta_{kq} = \frac{1}{L} \sum_j e^{-i(k-q)j}$ to simplify the above expression as
\begin{equation}
    H^{(0)} = \sum_{k,q} \delta_{kq} a^\ast_k a_q[2(\Delta + \Delta^{\prime}) - J(e^{iq} + e^{-ik})] = \sum_k [2(\Delta + \Delta^\prime) - 2J\cos{k}] n_k = \sum_k \epsilon_k n_k,
\end{equation}
where $\epsilon_k = 2(\Delta + \Delta^\prime) - 2J\cos{k}$. We have shown that, in the Fourier basis, $H^{(0)}$ is diagonal and so is indeed integrable. As discussed in the main paper, the term $H^{(2)}$ is non-linear and thus contributes to chaoticity of the system, which are parameterized by $\rho$.

\begin{figure}[!t]
    \includegraphics[width=0.5\columnwidth]{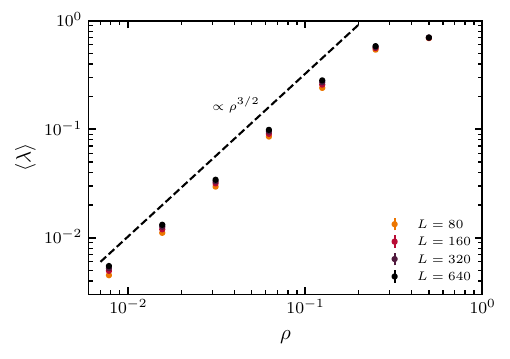}
	\caption{\textit{Lyapunov exponents of the classical XXZ model.} We consider the clean XXZ model with $W = 0$ at varying system sizes $L = 80, 160, 320, 640$. Then, we numerically compute the Lyapunov exponent for $2000$ initial spin configurations and plot their phase-space average $\langle \lambda \rangle$ as a function of magnetic density $\rho$ (and thus temperature $T$). We find that $\langle \lambda \rangle \propto \rho^{3/2}$ with the Lyapunov exponent decreasing as $\rho$ approaches zero. Parameters: $J = \sqrt{2}$, $\Delta = (\sqrt{5}+1)/4$, $\Delta^\prime = 1$, and $W = 0$ in $\HXXZ$.}
	\label{fig:clean_xxz_lyapunov}
\end{figure}

\section{Lyapunov exponents}\label{app:lyapunov}
The traditional measure of chaos in classical systems is performed using the so-called (maximum) Lyapunov exponent $\lambda$~\cite{Ott_2002}. The exponent $\lambda$ is computed by evolving two nearby trajectories $\vb{S}(t)$ and $\tilde{\vb{S}}(t)$ and measuring the distance $d(t) = \norm*{\vb{S}(t) - \tilde{\vb{S}}(t)}$ between them as a function of time $t$:
\begin{equation}
    \lambda = \lim_{t_\mathrm{evo}\to \infty}\lim_{d_0 \to 0} \frac{1}{t_\mathrm{evo}} \log{\left(\frac{d(t_\mathrm{evo})}{d_0}\right)},
    \label{eq:lyapunov}
\end{equation}
where $d_0 = d(t=0)$ is the initial distance and $t_\mathrm{evo}$ is the total evolution time. It is said that a system is chaotic if $\lambda > 0$ and integrable if $\lambda = 0$. In this section, we demonstrate that the $\HXXZ$ model (with $W = 0$) does indeed approach integrability as $\rho \to 0$ (or $T \to 0$) by numerically computing how $\lambda$ scales with $\rho$. This complements our results found in the main paper which shows this behavior using the Holstein-Primakoff representation and spectral functions. See also Ref.~\cite{ruidas2026manybodychaosemergespresence} for a related analysis in the Heisenberg model.

We numerically compute the Lyapunov exponents using the algorithm presented in Ref.~\cite{benettin1976kolmogorov} for varying system sizes $L = 80, 160, 320, 640$ with different values of magnetic density $\rho$ (and thus temperature $T$) in \autoref{fig:clean_xxz_lyapunov}. For each data point, we consider $2000$ initial spin configurations, compute the Lyapunov exponent for each trajectory, and calculate the phase-space averaged Lyapunov exponent $\langle \lambda \rangle$. Our results in \autoref{fig:clean_xxz_lyapunov} verify that the Lyapunov exponent decreases as $\rho \to 0$ ($T \to 0$), confirming our findings that $\rho \to 0$ ($T \to 0$) is indeed an integrable limit. 

In addition, we find that the Lyapunov time $T_\mathrm{lya}$ scales as $T_\mathrm{lya} \equiv 1 / \expval{\lambda} \propto \rho^{-3/2}$ while, in \figref{fig:clean_xxz_current}{a}, we find that the relaxation/thermalization time $T_\mathrm{th}$ scales as $T_\mathrm{th} \propto \rho^{-2}$. This suggests that $T_\mathrm{lya}$ is asymptotically faster than $T_\mathrm{th}$ when $\rho \to 0$ and indicates that the classical XXZ model $\HXXZ$ at $W = 0$ exhibits confined chaos~\cite{milani_an_1992,milani_stable_1997,laskar_chaotic_2008,winter_short_2010,goldfriend2019equilibration,goldfriend2020quasi,pereira_confined_2024,kim2025confineddeconfinedchaosclassical}, where the exponential divergence of trajectories occurs well before relaxation and thermalization.

\begin{figure}[!b]
    \includegraphics[width=\columnwidth]{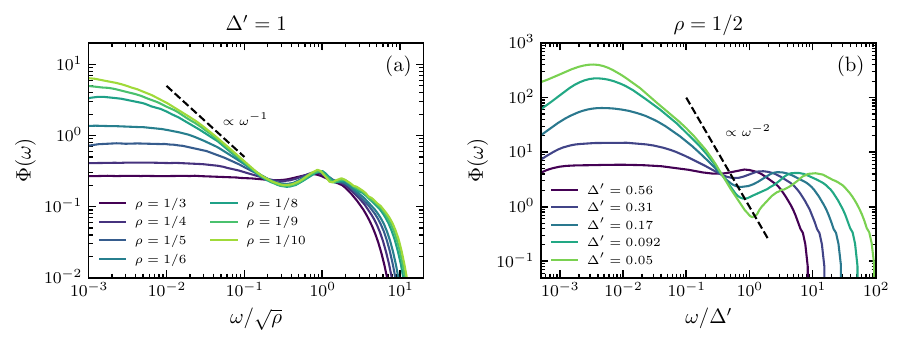}
	\caption{\textit{Spectral functions of the disorder-less quantum model.} The spectral functions $\Phi(\omega)$ develop low-frequency power law tails near the integrable points. (a): At low temperatures (small $\rho$'s), $\Phi(\omega) \propto \omega^{-1}$. (b): At weak integrability-breaking strengths, $\Phi(\omega) \propto \omega^{-2}$. Parameters: $J=\sqrt{2}$, $\Delta=(\sqrt{5}+1)/4$, and $W = 0$ in $\HXXZ$ at a fixed semi-momentum sector with $k = (2\pi/L) \left\lfloor L/4 \right\rfloor$ for $L \in [22, 50]$ in (a) and $L = 22$ in (b). For all the lines, we average over the central $75\%$ of eigenstates.}
	\label{fig:clean_xxz_spec}
\end{figure}

\section{Spectral functions in the disorder-less quantum model}\label{app:clean_spec}

In this section, we analyze the spectral function corresponding to the disorder-less quantum model with $W = 0$ in $\HXXZ$, which is consistent with the quantum model considered in Fig.~2 of the main text. As in the main text, we choose the probe operator $O=\partial_{\Delta} H_{\rm XXZ}$ and plot its spectral function $\Phi(\omega)$ at low temperature and strong integrability-breaking strength (see \figref{fig:clean_xxz_spec}{a}), and at high temperature and weak integrability-breaking strength (see \figref{fig:clean_xxz_spec}{b}). At low temperatures and strong integrability-breaking strengths, the spectral function $\Phi(\omega) \propto \sqrt{\rho}/\omega$ at low frequencies, indicating a logarithmically slow long-time relaxation of the the observable $O$. On the other hand, at high temperatures and weak integrability-breaking strengths, the spectral function scales as $\Phi(\omega) \propto \Delta^{\prime 2}/\omega^{2}$ at low frequencies, which is consistent with an exponential relaxation as predicted by Fermi's Golden Rule (FGR). This latter scaling for the XXZ model has also found been in Ref.~\cite{leblond2021universality}.

\section{Spectral functions in a quantum ladder model}\label{app:ladder}

We extend our analysis of the quantum model by considering the XXZ ladder. Here, we compute spectral functions as we vary either the magnetic density (temperature) or the integrability-breaking perturbation strength. The Hamiltonian of the interaction-coupled XXZ ladder model is given as
\begin{equation} \label{eq:quantum_ladder_model}
    H_\text{ladder} = -\left[ \frac{J}{2}\sum^{L-1}_{j=0}\left(\hat{S}^{+}_{j} \hat{S}^{-}_{j+2} + \mathrm{h.c.} \right) + \Delta\sum^{L-1}_{j=0} \hat{S}^z_j \hat{S}^z_{j+2} + U\sum^{L-1}_{j=0} \hat{S}^z_j \hat{S}^z_{j+1} \right], 
\end{equation}
such that even and odd sites represent two different legs of the ladder and $L$ is even. $J$ and $\Delta$ are hopping and nearest-neighbor interaction strengths along each leg, respectively. $U$ is the nearest-neighbor interaction strength connecting the two legs. For $U=0$, we recover two decoupled integrable XXZ spin chains along odd and even sites. The interaction $U$ breaks integrability as it leads to interactions between legs of the ladder. The system therefore approaches an integrable point with either decreasing $T$ or $U$. For our probe operator, we choose the total kinetic energy ($\mathrm{KE}$) $O=\partial_{J}H_{\rm ladder}$, and calculate the spectral function $\Phi_\mathrm{KE}(\omega)$.

\begin{figure}[!b]
    \includegraphics[width=\columnwidth]{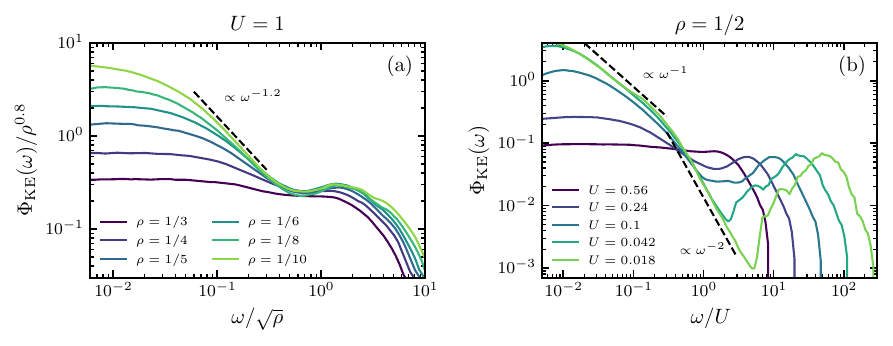}
	\caption{\textit{Spectral function in the interaction-coupled quantum XXZ ladder model.} The spectral function scales as $\Phi_\mathrm{KE}(\omega)\propto \omega^{-1.2}$ and $\Phi_\mathrm{KE}(\omega)\propto \omega^{-1}$ at low frequencies near the integrable point for (a) low temperature and high integrability-breaking strength and (b) high temperature and low integrability-breaking strength, respectively. In (b), at intermediate frequencies, $\Phi(\omega)$ also shows a $\omega^{-2}$ tail which indicates that $\Phi(\omega) \propto (U/\omega)^2$ due to the data collapse. Parameters: $J=\sqrt{2}$ and $\Delta=(\sqrt{5}+1)/4$ in $H_\mathrm{ladder}$ (\autoref{eq:quantum_ladder_model}) at the semi-momentum sector $k = (2\pi/L) \left\lfloor L/4 \right\rfloor$ for $L \in [22, 50]$ in (a) and $L = 22$ in (b).For all the lines, we average over the central $75\%$ of eigenstates.}
	\label{fig:ladder_XXZ_spec1}
\end{figure}

\begin{figure}[!htbp]
    \includegraphics[width=0.5\columnwidth]{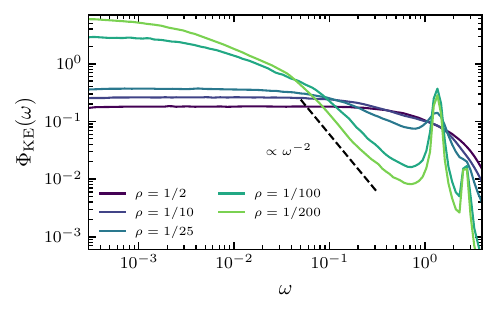}
	\caption{\textit{Spectral function in the fully-coupled quantum XXZ ladder model.} The spectral function scales as $\Phi_{\rm KE}(\omega) \propto \omega^{-2}$ near the integrable point as we lower magnetic density $\rho$ (and thus temperature $T$). Parameters: $J=\sqrt{2}$ and $\Delta=(\sqrt{5}+1)/4$ in $H_\mathrm{ladder}^{\rm full}$  (\autoref{eq:quantum_fully_ladder_model}) at the semi-momentum sector $k = (2\pi/L) \left\lfloor L/4 \right\rfloor$ for $L \in [22, 600]$. For all the lines, we average over central $75\%$ of states.}
	\label{fig:ladder_XXZ_spec2}
\end{figure}

We observe a low-frequency scaling $\Phi_\mathrm{KE}(\omega) \propto \rho^{1.4}\omega^{-1.2}$ when we fix $U = 1$ and decrease $\rho$  in \figref{fig:ladder_XXZ_spec1}{a} (which is close to $\omega^{-1}$ as seen in the one-dimensional XXZ model [see \figref{fig:clean_xxz_spec}{a}]), and $\Phi_\mathrm{KE}(\omega) \propto (U/\omega)^2$ when we fix $\rho = 1/2$ and decrease $U$ in \figref{fig:ladder_XXZ_spec1}{b}.

We now consider the fully-coupled quantum XXZ ladder Hamiltonian in which particles can also hop between the two legs of the ladder in addition to the nearest-neighbor interaction:
\begin{equation} \label{eq:quantum_fully_ladder_model}
    H_\mathrm{ladder}^{\rm full} = -\left[\frac{J}{2}\sum^{L-1}_{j=0}\left(\hat{S}^{+}_{j}\hat{S}^{-}_{j+2}+ \mathrm{h.c.} \right) + \frac{J}{2}\sum^{L-1}_{j=0}\left(\hat{S}^{+}_{j}\hat{S}^{-}_{j+1}+ \mathrm{h.c.} \right) +  \Delta\sum^{L-1}_{j=0} \hat{S}^z_j\hat{S}^z_{j+1}+\Delta\sum^{L-1}_{j=0}\hat{S}^z_j \hat{S}^z_{j+2} \right].
\end{equation}
We then study the relaxation of the kinetic energy operator $O = \partial_{J} H_\mathrm{ladder}^{\rm full}$ by calculating its spectral function. In this model, we instead recover the scaling $\Phi_{\rm KE}(\omega) \propto \omega^{-2}$ at low frequencies as we decrease $\rho$ in \autoref{fig:ladder_XXZ_spec2}, which is similar to the classical one-dimensional model and consistent with FGR.

\section{Thouless frequency and Drude weight}\label{app:thouless-scaling}

\begin{figure}[!t]
    \includegraphics[width=\columnwidth]{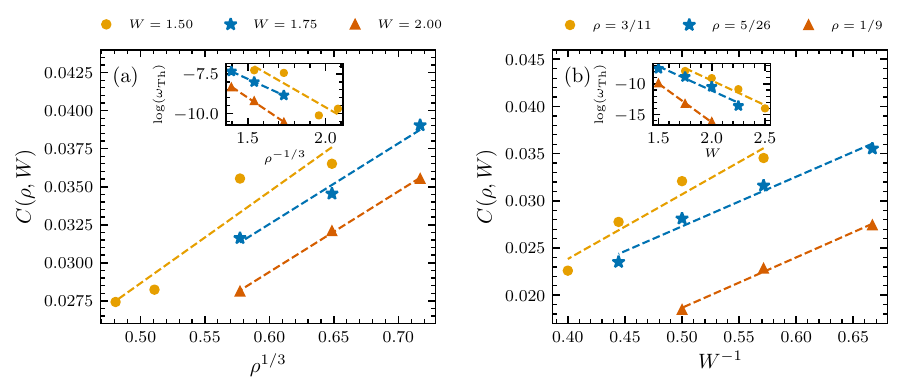}
	\caption{\textit{Thouless frequency extraction in the disordered quantum model.} We extract the slope $C=C(\rho,W)$ of the spectral function $\Phi(\omega) \propto C/\omega$. (a): $C(\rho,W)$ at fixed $W$ scales as $C(\rho,W) \propto \rho^{1/3}$. (b): $C(\rho,W)$ at fixed $\rho$ scales as $C(\rho,W) \propto W^{-1}$. For both (a) and (b), the dashed lines are linear fits. Insets: Thouless frequency $\omega_\mathrm{Th}$ is computed using \autoref{eq:sm_thouless}: $\omega_\mathrm{Th} \propto \mathrm{exp}(-\rho^{-1/3})$ in (a) and $\omega_\mathrm{Th} \propto \exp(-W)$ in (b).
    Parameters: $J=2.0$, $\Delta=1.0$ and $\Delta^{\prime}=1.5$ in $H_\mathrm{XXZ}$ for $L \in [18, 36]$.}
	\label{fig:thouless_sm}
\end{figure}

In this section, we estimate the Thouless frequency by extrapolating the low-frequency power-law tail of the spectral function for the disordered quantum model $\HXXZ$. As shown in the main text, near the onset of localization, $\Phi(\omega)=C\omega^{-1}$ at fixed disorder $W$ while varying magnetic density $\rho$ and at fixed $\rho$ while varying disorder strength $W$ (see Figs.~4a and 4b of the main text, respectively). However, as argued in the main text, if we extrapolate this scaling towards the thermodynamic limit (as $L \to \infty$ and thus $\omega_\mathrm{H} \to 0$) while fixing $\rho$ and $W$, then there must exist a frequency where it stops or crossovers into a slower scaling regime to preserve normalization of the spectral function. We identify this point as the Thouless frequency, which is inverse of the relaxation time or Thouless time.

Here, we analytically calculate the Thouless frequency using the sum rule for the spectral function (see also Ref.~\cite{sels2020dynamical} for a similar analysis). First, we numerically extract the slope $C$ by plotting $\omega\Phi(\omega)$ against $\omega$ for each data-set corresponding to different pairs of $(\rho, W)$. $\omega\Phi(\omega)$ shows a plateau where $\Phi(\omega) \propto C \omega^{-1}$ and has a point of inflection $\omega_\mathrm{in}$ where the sign of its curvature changes. We numerically detect the height of this plataeu as $C \equiv \omega_\mathrm{in} \Phi(\omega_\mathrm{in})$ at this inflection point. Further, from Figs.~4a and 4b of the main text, the low-frequency spectral tail $\omega^{-1}$ collapses when scaling the $x$-axis as $\rho^{-1/3}\omega$ at fixed $W$ and as $W\omega$ at fixed $\rho$, respectively. We denote the onset frequency of the $\omega^{-1}$ scaling as the ultraviolet (UV) frequency $\omega_\mathrm{uv}$ (located in the high-frequency region). From the scaling collapses, we define $\omega_{\mathrm{uv}}$ to be at $\rho^{-1/3} \omega_\mathrm{uv} = 1$ at fixed $W$ and $W \omega_\mathrm{uv} = 1$ at fixed $\rho$, respectively.

Since $\omega_{\rm uv}$ drifts linearly  with either $\rho^{1/3}$ or $W^{-1}$, in the asymptotic scaling limit, almost all of the spectral weight is dominated by the low-frequency $\omega^{-1}$ tail as we approach the transition between chaotic and integrable regimes. From the sum rule, $2\int^\infty_0 \Phi(\omega) \dd \omega = 1$, we equate $\int_{\omega_{\rm Th}}^{\omega_{\rm uv}}\Phi(\omega) \dd \omega = \tfrac{1}{2} - I_\mathrm{uv}$, where $I_\mathrm{uv} = \int^\infty_{\omega_\mathrm{uv}} \Phi(\omega) \dd \omega$ is the contribution of the high-frequency spectral weight. This gives us an estimate of the Thouless frequency:
\begin{equation} \label{eq:sm_thouless}
    \int_{\omega_{\rm Th}}^{\omega_{\rm uv}}\frac{C}{\omega} \dd \omega =\frac{1}{2}- I_\mathrm{uv}\implies\omega_{\rm Th}  = \omega_{\rm uv} \exp(-\frac{1}{2C} + \frac{I_\mathrm{uv}}{C}).
\end{equation}
Here, we make the implicit assumption that the Drude weight supplements the $\omega^{-1}$ tail contribution up to $\omega_\mathrm{Th}$ in the thermodynamic limit. Note that the observable $O = \partial_{\Delta}\HXXZ$ has a finite Drude weight in an infinite temperature ensemble at fixed magnetization, even in the thermodynamic limit. This can be seen analytically by calculating $\langle O\rangle = \Omega^{-1}{\rm Tr}_{M}\left[\sum\limits_{j=0}^{L-1}\hat{S}_j^z \hat{S}_{j+1}^z\right] \mathbbm{1}$ under periodic boundary conditions. Only the states that satisfy the following constraints will contribute to this trace:  nearest-neighbor spins are both aligned, (i) $\expval*{\hat{S}_j^z}=\expval*{\hat{S}_{j+1}^z}=1/2$, (ii) $\expval*{\hat{S}_j^z}=\expval*{\hat{S}_{j+1}^z}=-1/2$, and nearest-neighbor spins are oppositely aligned, (iii) $\expval*{\hat{S}_j^z}=1/2$, $\expval*{\hat{S}_{j+1}^z}=-1/2$, (iv) $\expval*{\hat{S}_j^z}=-1/2$, $\expval*{\hat{S}_{j+1}^z}=1/2$. Given the condition $M=S^z_{\rm tot}$, the number of states contributing to each of these situations are $\begin{pmatrix}L-2\\ \rho L-2\end{pmatrix}$, $\begin{pmatrix}L-2\\ \rho L\end{pmatrix}$, $\begin{pmatrix}L-2\\ \rho L-1\end{pmatrix}$ and $\begin{pmatrix}L-2\\ \rho L-1\end{pmatrix}$, respectively. Therefore,
\begin{equation}
    \Omega^{-1}{\rm Tr}_{M}\left[\sum\limits_{j=0}^{L-1}\hat{S}_j^z \hat{S}_{j+1}^z\right] = \frac{L\Omega^{-1}}{4}\left[\begin{pmatrix}L-2\\ \rho L-2\end{pmatrix} + \begin{pmatrix}L-2\\ \rho L\end{pmatrix} - 2 \begin{pmatrix}L-2\\ \rho L-1\end{pmatrix}\right] = \frac{L(2\rho - 1)^2}{4} + \frac{L\rho(\rho-1)}{L-1}.
\end{equation}
This part of the Drude weight will not contribute to the Thouless frequency as we extrapolate the $\omega^{-1}$ tail of the spectral function to the thermodynamic limit. For open boundary conditions, one would further have $\mathcal{O}(1)$ corrections to the Drude weight. To discard this conserved weight, we numerically compute $\expval{O}$ for each $L$ and $\rho$ in our simulations and identify the traceless observable $O\rightarrow O-\langle O \rangle$ as done everywhere in the paper. 
As we approach the thermodynamic limit, any finite-size Drude weight of traceless $O$ must therefore vanish. For reference, with the system sizes available to us, at a fixed $W = 1.5$ in \figref{fig:thouless_sm}{a}, the Drude weight contribution remains around $7\%$ as $\rho$ decreases while $I_\mathrm{uv}$ remains around $25\%$. On the other hand, at a fixed $\rho = 3/11$ in \figref{fig:thouless_sm}{b}, the Drude weight contribution increases from $7\%$ to $17\%$ of $\norm{O}^2$ as $W$ increases while $I_\mathrm{uv}$ decreases from $25\%$ to $20\%$. 

\begin{figure}[!t]
    \includegraphics[width=0.5\columnwidth]{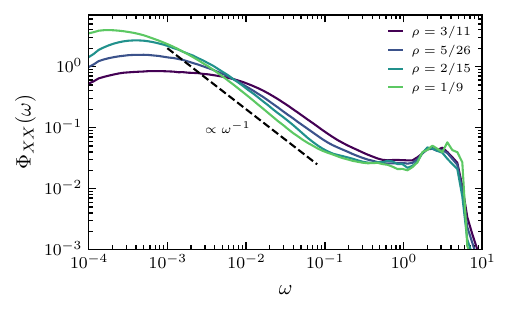}
	\caption{\textit{Spectral function of $XX$ in the disordered quantum model.} Following the same parameters used in Fig.~4a of the main text, we compute the spectral function of $XX = \sum_j \hat{S}_j^x \hat{S}_{j+1}^x$ in $\HXXZ$ for different values of $\rho$. We find that $\Phi_{XX} \propto \omega^{-1}$ as $\omega \to 0$.  Parameters: $J = 2$, $\Delta = 1$, $\Delta^\prime = 1.5$, and $W = 1.5$ in $\HXXZ$.}
	\label{fig:XX_spec}
\end{figure}

In \figref{fig:thouless_sm}{a}, we plot $C= C(\rho,W)$ as a function of $\rho^{1/3}$ for three values of $W$, which indicates that, at least in the regime of our chosen parameters, $C\propto\rho^{1/3}$ for a fixed $W$. This is consistent with the scaling collapse of the spectral function in Fig.~4a of the main text since $\Phi(\omega) \propto \rho^{1/3}/\omega = (\omega / \rho^{1/3})^{-1}$. In \figref{fig:thouless_sm}{b}, we plot $C(\rho, W)$ as a function of $W^{-1}$ for three different values of $\rho$ and obtain the scaling $C\propto W^{-1}$ at fixed $\rho$, which is consistent with Fig.~4b of the main text. In the insets of \figref{fig:thouless_sm}{a} and \figref{fig:thouless_sm}{b}, we plot the calculated Thouless frequency using \autoref{eq:sm_thouless} against $\rho^{-1/3}$ at fixed $W$ and against $W$ at fixed $\rho$, respectively. Collectively, they demonstrate that, as we approach the transition, the Thouless frequency scales as $\omega_{\rm Th}\propto \exp(-\rho^{-1/3})$ at fixed $W$ and $\omega_{\rm Th} \propto \exp(-W)$ at fixed $\rho$.

In addition, we have checked the relaxation of other observables such as $XX \equiv \sum_j\hat{S}^x_j\hat{S}^x_{j+1}$ (see \autoref{fig:XX_spec}) and observe qualitatively similar low-frequency power-law tails in the spectral function as $\rho$ decreases. This indicates that approaching an integrable point by decreasing temperature is a generic effect. However, for $XX$, the Drude weight takes up a significant part of $\norm{XX}^2$: Drude weight contributions range between $51\%$ and $57 \%$ of $\norm{XX}^2$. We numerically confirmed that this is due to a large dependence of $\mel{n}{XX}{n}$ on energies, leading to non-zero values of $\mel{n}{XX}{n} - \Omega^{-1} \Tr[XX] = \mel{n}{XX}{n}$ (since $\Tr[XX] = 0$) as one moves away from the center of the spectrum. 
 
\begin{figure}[!b]
    \includegraphics[width=\columnwidth]{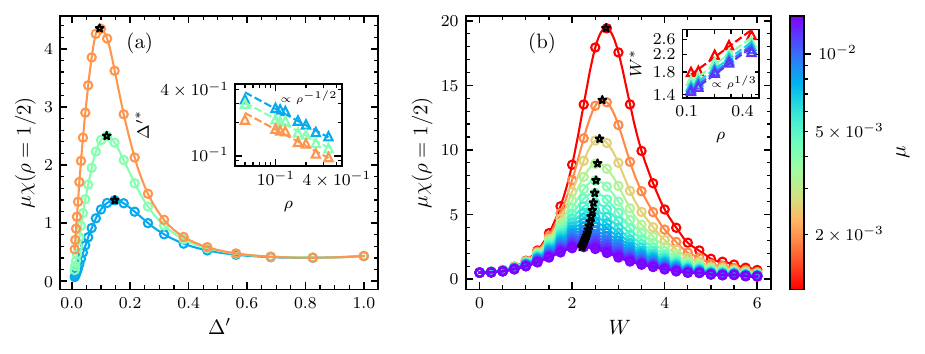}
	\caption{\textit{Fidelity susceptibility in the disorder-less and disordered quantum models.} We compute the fidelity susceptibility $\chi$ for varying $\Delta^\prime$ with $W = 0$ in (a) and varying $W$ with $\Delta^\prime=1.5$ in (b) at different values of $\mu>\omega_H$ for the Hamiltonian $\HXXZ$. The fidelity susceptibility peak drifts to lower $\Delta^\prime$ or higher $W$ with decreasing values of $\mu$ at fixed magnetic density $\rho = 1/2$. Insets: The drifts of the peaks with magnetic density $\rho$ (and thus temperature $T$) are given by ${\Delta^\prime}^\ast \propto \rho^{-1/2}$ in (a) and $W^\ast \propto \rho^{1/3}$ in (b). In (a), we consider the semi-momentum sector $k = (2\pi/L) \left\lfloor L/4 \right\rfloor$ for $L = 22$, while, in (b), we set $L = 18$.}
	\label{fig:chi_drift_quantum}
\end{figure}

\section{Drift scaling of fidelity susceptibility with temperature}\label{app:drift-scaling}
We analyze the drift of the fidelity susceptibility maximum with temperature in the one-dimensional quantum model $\HXXZ$. In \figref{fig:chi_drift_quantum}{a} and \figref{fig:chi_drift_quantum}{b}, we compute the fidelity susceptibility $\chi$ at fixed magnetic densities $\rho$ while varying either $\Delta^\prime$ in the disorder-less quantum model ($W = 0$) or $W$ in the disordered quantum model ($\Delta^\prime = 1.5$), respectively. As $\mu \to 0$, the peak of the fidelity susceptibility drifts towards either $\Delta^\prime \to 0$ or $W^{-1} \to 0$. For the disordered quantum model, we argue in the main text that this drift indicates exponentially long prethermalization times in $W$, which is consistent with earlier studies.

On the other hand, at a fixed $\mu>\omega_H$ with varying $\rho$, we see algebraic drifts of the fidelity susceptibility peaks with $\Delta^{\prime}={\Delta^\prime}^\ast \propto \rho^{-1/2}$ in \figref{fig:chi_drift_quantum}{a} and $W=W^\ast \propto \rho^{1/3}$ in 
\figref{fig:chi_drift_quantum}{b}. The latter scaling is consistent with $W^\ast \propto -\rho^{1/3} \log(\mu)$ as reported in the main text (see Fig.~3b in main text). 

\section{Relaxation of spin current: transport}\label{app:current}

Following Kubo linear response theory \cite{toda2012statistical}, the auto-correlation function of the spin current is directly proportional to the spin conductivity in the system. Enhanced low-frequency spectral weight in its Fourier transform indicate anomalously slow spin transport. This section showcases how the total spin current density relaxes to thermal equilibrium in both quantum and classical models near either an integrable point or at low temperatures. 

\begin{figure}[!b]
    \includegraphics[width=\columnwidth]{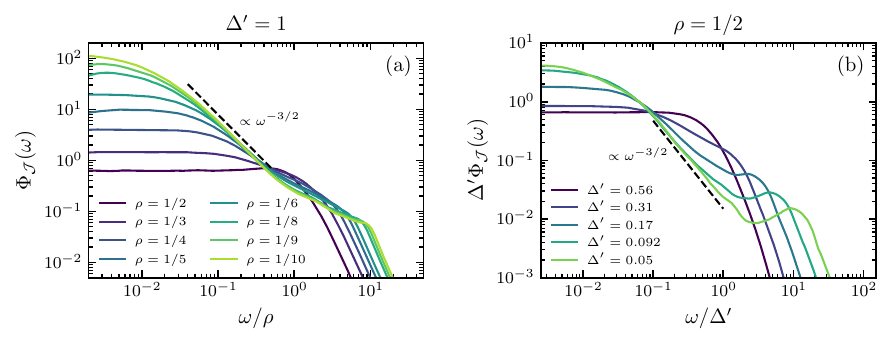}
	\caption{\textit{Relaxation of the spin current density $\mathcal{J}$ in the disorder-less quantum model.} The spectral function $\Phi_\mathcal{J}(\omega)$ develops low-frequency power law tails near the integrable points at (a) low temperature and strong integrabality-breaking strength and (b) weak integrability-breaking strength and high temperature. For both cases, the spectral function scales as $\Phi_\mathcal{J}(\omega) \propto \omega^{-3/2}$. Parameters: $J=\sqrt{2}$, $\Delta=(\sqrt{5}+1)/4$, and $W = 0$ in $\HXXZ$ at the semi-momentum sector $k = (2\pi/L) \left\lfloor L/4 \right\rfloor$ for $L \in [22, 50]$ in (a) and $L = 22$ in (b). For all the lines, we average over the central $75\%$ of eigenstates.}
	\label{fig:quantum_XXZ_current}
\end{figure}

\begin{figure}[!t]
    \includegraphics[width=\columnwidth]{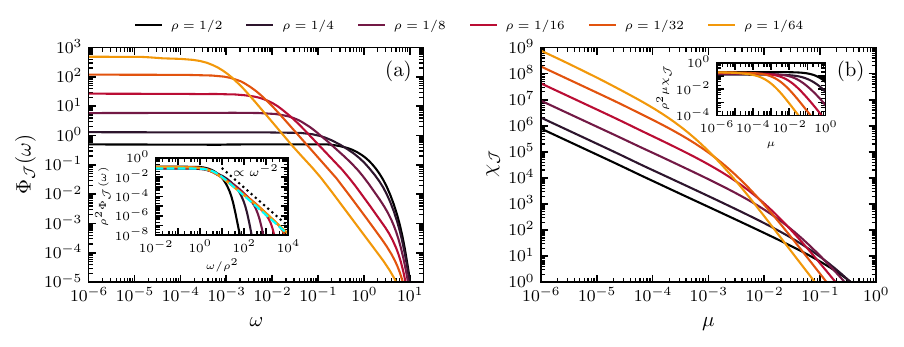}
	\caption{\textit{Relaxation of the spin current density $\mathcal{J}$ in the disorder-less classical model.} We plot the relaxation of $\mathcal{J}$ at varying values of magnetic density $\rho$ and thus temperature $T$. (a) The spectral function develops a low-frequency tail with scaling $\Phi_\mathcal{J}(\omega) \propto \rho^4/\omega^2$. We obtain scaling collapse for $\rho^2 \Phi_\mathcal{J}(\omega)$ against $\omega/\rho^2$ in the inset, where the blue dashed line indicates a Lorentzian. (b) The fidelity susceptibility $\chi_\mathcal{J} \propto 1/\mu$ as $\mu \to 0$. In the inset, we obtain collapse for $\rho^2 \mu \chi_\mathcal{J}$ against $\mu$.  Parameters: $J = \sqrt{2}$, $\Delta = (\sqrt{5}+1)/4$, $\Delta^\prime = 1$, $W = 0$ in $\HXXZ$, and $L = 640$.}
	\label{fig:clean_xxz_current}
\end{figure}

In the one-dimensional disorder-less XXZ models $\HXXZ$ (with $W = 0$), the total spin current density $\hat{\mathcal{J}} = \frac{1}{L} \sum_\ell \hat{j}_\ell$ is defined using the identities
\begin{equation}
    \pdv{\hat{S}_\ell^z}{t} = -i \comm{\hat{S}_\ell^z}{\HXXZ},~\pdv{S_\ell^z}{t} = \poissonbracket{S^z_\ell}{H_\mathrm{XXZ}},
\end{equation}
where the equation on the left-hand side (right-hand side) corresponds to the quantum (classical) XXZ model. In both cases, we have that
\begin{equation}
    \pdv{\hat{S}_\ell^z}{t} = \hat{j}_\ell - \hat{j}_{\ell+1},~ \pdv{S_\ell^z}{t} = j_\ell - j_{\ell+1},
\end{equation}
where $\hat{j}_\ell = J(\hat{S}^x_\ell \hat{S}^y_{\ell+1} - \hat{S}^y_\ell \hat{S}^x_{\ell+1})$ and $j_\ell = J(S^x_\ell S^y_{\ell+1} - S^y_\ell S^x_{\ell+1})$. Consequently, the total spin current density $\mathcal{J}$ can be computed by averaging over all sites.

Focusing on the quantum model with $W = 0$, we compute the spectral function of $\mathcal{J}$, $\Phi_\mathcal{J}(\omega)$, either by fixing $\Delta^\prime = 1$ and varying $\rho$ in \figref{fig:quantum_XXZ_current}{a} or by fixing $\rho = 1/2$ and varying $\Delta^\prime$ in \figref{fig:quantum_XXZ_current}{b}. In both scenarios, we observe emerging low-frequency tails of the spectral function as we approach their respective integrable limits with scaling relations $\Phi_\mathcal{J}(\omega) \propto (\rho/\omega)^{3/2}$ and $\Delta^{\prime}\Phi_\mathcal{J}(\omega) \propto (\Delta^\prime/\omega)^{3/2}$, respectively. These suggest power-law relaxations of the auto-correlation functions of $\mathcal{J}$.

On the other hand, in the classical model, we perform a similar calculation by computing $\Phi_\mathcal{J}(\omega)$. However, as also discussed in the main text, this model is not integrable even if $\Delta^\prime = 0$, and thus we only fix $\Delta^\prime = 1$ and vary $\rho \to 0$ to reach the integrable limit. In \figref{fig:clean_xxz_current}{a}, the spectral function scales as $\Phi_\mathcal{J}(\omega) \propto \rho^4/\omega^2$, where we observe a scaling collapse by re-scaling $\rho^2 \Phi_\mathcal{J}(\omega)$ against $\omega/\rho^2$ as shown in the inset of \figref{fig:clean_xxz_current}{a}. Here, the spectral function follows a Lorentzian $\Phi_\mathcal{J}(\omega) \approx \frac{1}{\pi} \frac{\Gamma}{\Gamma^2 + \omega^2}$, where $\Gamma$ is the Lorentzian width, in agreement with the FGR prediction. This implies that the thermalization time $T_\mathrm{th}$ of the current density scales as $T_\mathrm{th} \equiv 2\pi/\Gamma \propto \rho^{-2}$. In \figref{fig:clean_xxz_current}{b}, we provide an alternative perspective by computing the fidelity susceptibility $\chi_\mathcal{J}$ at varying values of $\mu$ using the spectral functions from \figref{fig:clean_xxz_current}{a} and Eq.~10 of the main text. In agreement with the Lorentzian functional form of the spectral function, we obtain a scaling collapse for $\rho^2 \mu \chi_\mathcal{J}$ against $\mu$ in the inset.

\bibliography{refs}